\pdfoutput=1

\documentclass[11pt]{article}

\usepackage{acl}

\usepackage{times}
\usepackage{latexsym}
\usepackage{graphicx}
\usepackage{multirow}

\usepackage{booktabs}
\usepackage{xltabular}
\usepackage[T1]{fontenc}

\usepackage[utf8]{inputenc}

\usepackage{microtype}

\title{Adversarial Examples for Evaluating Conversational Recommender Systems}
\title{Evaluating the Robustness of Conversational Recommender Systems by Adversarial Examples}

 \author{Ali Montazeralghaem \and James Allan \\
       Center for Intelligent Information Retrieval \\ Manning College of Information and Computer Sciences \\ University of Massachusetts Amherst}

\begin{document}
\maketitle
\begin{abstract}
Conversational recommender systems (CRSs) are improving rapidly, according to the standard recommendation accuracy metrics. However, it is essential to make sure that these systems are \emph{robust} in interacting with users including regular and malicious users who want to attack the system by feeding the system modified input data. In this paper, we propose an adversarial evaluation scheme including four scenarios in two categories and automatically generate adversarial examples to evaluate the robustness of these systems in the face of different input data. By executing these adversarial examples we can compare the ability of different conversational recommender systems to satisfy the user's preferences. We evaluate three CRSs by the proposed adversarial examples on two datasets. Our results show that none of these systems are robust and reliable to the adversarial examples. 
\end{abstract}

\section{Introduction}

\label{sec:intro}
A Conversational Recommender System (CRS) is a task-oriented dialogue system that guides users through a multi-turn conversational interaction to fulfill recommendation-related objectives \cite{sun2018conversational, zhang2018towards, montazeralghaem2021large}. 
In each turn of the conversation, the system can ask a question to clarify user needs and infer the dynamic preferences of users from their utterances, and hopefully, the system can provide the right items to the user when she is fully confident of the user needs. 
Conversational recommender systems consist of four components: Natural Language Understanding (NLU), Dialog State Tracker (DST), Dialog Policy, and Natural Language Generation (NLG) \cite{zhou2020improving, li2018towards}. Most of the existing methods developed NLU and NLG modules by using the encoder-decoder framework and different machine learning techniques for two other components (i.e., DST and Dialog Policy).

Due to the large range of system evaluation settings and configurations, evaluating and comparing the overall performance of conversational recommender systems is particularly challenging. Several techniques have been proposed for different components in these systems, whereas these components are merely evaluated separately. Most existing methods compare their models with baselines separately on different components, assuming that a successful conversational recommender system can be built by assembling a set of successful components.
However, a component can be better than baselines individually, but its impact on the whole system is not clear, and few studies investigated the overall performance of these components. 

In this paper, we focus on the impact of the NLU component on conversational recommender systems. Most of the existing models appear relatively successful by standard evaluation metrics in recommending items in the conversational recommendation paradigm. 
However, existing methods do not seem to have the true ability to satisfy user preferences in a conversation and reasoning on them.  In this paper, we propose adversarial evaluation for conversational recommender systems, by  evaluating language understanding of these systems on adversarially-chosen inputs. 

Adversarial evaluation of conversational recommender systems needs new methods to assess the ability of these systems to understand the needs of users in conversations. Prior works in computer vision and natural language processing have shown that small imperceptible adversarial perturbations to input (i.e., images or text) can change the true label of the input \cite{szegedy2013intriguing, goodfellow2014explaining}. On the other hand, in a conversation between a recommender system and a user, we can distract the system with adversarial answers to system questions by altering the user's utterances. We propose four scenarios in two categories to generate adversarial answers for conversational recommender systems to evaluate them in this way. 

In the proposed scenarios, we alter the user's answer in a turn of the conversation to evaluate the output of the system. These scenarios are divided into two different categories. In the first category, by \emph{changing or adding more details} to the user's answer, we expect  CRS models generate \emph{same} predictions. In contrast, In the next category, we expect different predictions by \emph{changing or adding a contradictory sentence} to the user's answer.

By executing these evaluations, we can compare the ability of different conversational recommender systems to satisfy user preferences, and as a result, show the reliability of these systems.

Our experiments are performed on CRSLab \cite{zhou2021crslab} which is an open-source toolkit for building conversational recommender systems. We evaluate three CRS models on two datasets and show that none of them are robust and reliable to the adversarial examples. Our experiments show that CRS systems need to be more sensitive to user responses, which leads users to have more confidence in these systems. We have made all of our code and data publicly available to stimulate the creation of new models that comprehend language more precisely in conversational recommender systems.

\begin{figure}
  \centering
    \includegraphics[width=0.45\textwidth]{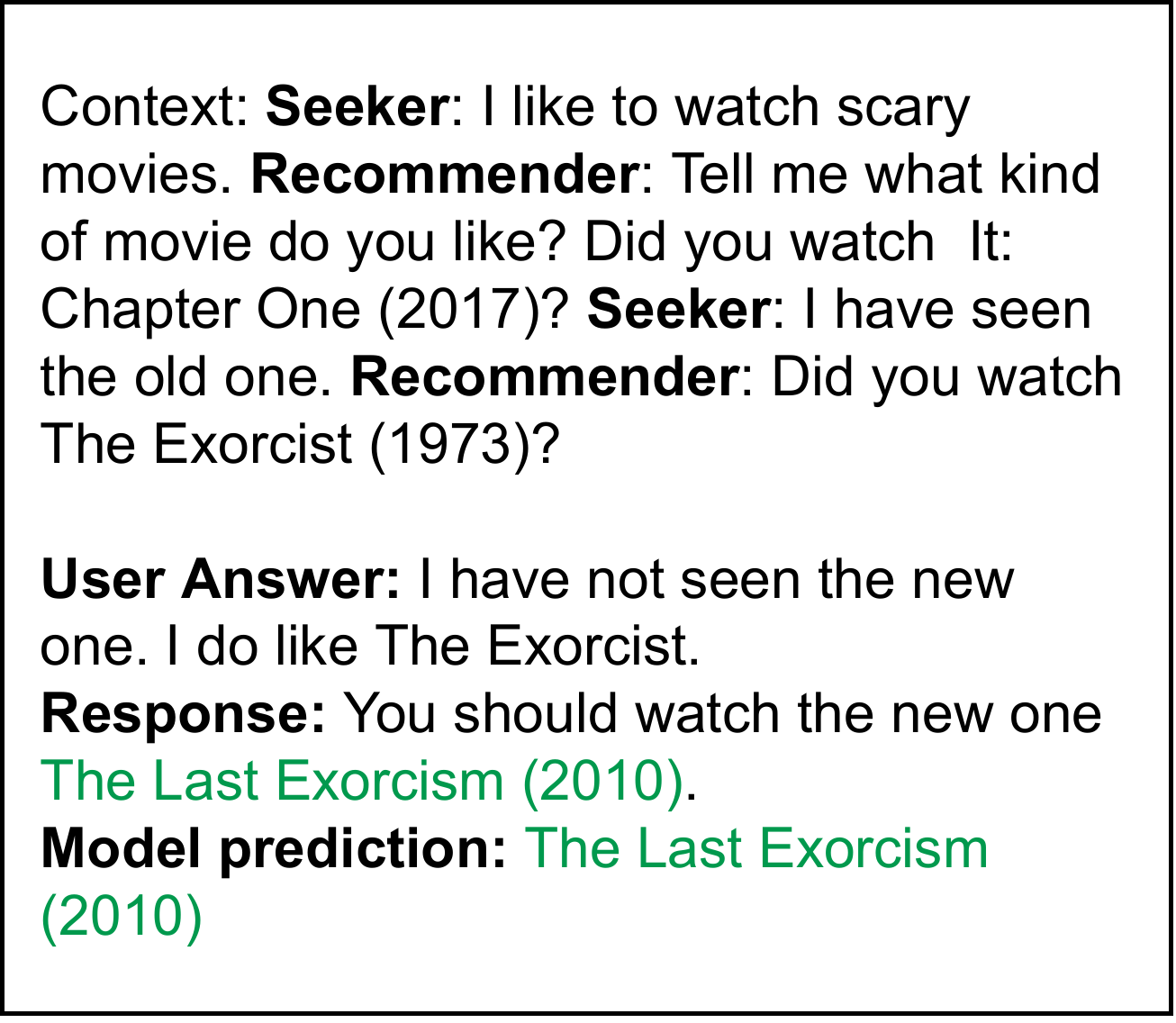}
    \caption{Original Conversation between Seeker (i.e, user) and Recommender.}
    \label{fig:orginal_adv}
\end{figure}

\section{Conversational Recommender Systems}
Recently, with progress in natural language processing (NPL) conversational recommender systems have achieved much more attention \cite{chen2019towards, li2018towards, liao2019deep, sun2018conversational, lei2020estimation, montazeralghaem2022extracting, montazeralghaem2022learning}. Unlike standard recommender systems, conversational recommender systems can provide high-quality recommendations through conversations with users. In other words, a system in this area can ask clarification questions in natural languages through the conversation and give user feedback. When the system gets enough information and is confident from the user's needs, it recommends some items to the user. 

In this paper, we want to evaluate the robustness of conversational recommender systems. In general, we are given the context of the conversation $C$ and the user's last answer $A$ and we aim to convert the $A$ to $A'$ to fool the system. Note that we know what is the response of a conversational recommender system $CRS$ to the user's last answer i.e., $R = CRS(C, A)$. An example of an original conversation between a seeker and a recommender system is shown in Figure \ref{fig:orginal_adv}.

\section{Adversarial Evaluation}
To fool a conversational recommender system, we design an adversarial evaluation scheme that includes four scenarios in two categories: 

\begin{itemize}
    \item \textbf{Cat1} expecting the \emph{same} prediction by \emph{changing} the user's answer or \emph{adding more details} to the user's answer, and
    
    \item \textbf{Cat2} expecting a \emph{different} prediction by \emph{changing} the user's answer or \emph{adding a contradictory sentence} to the user's answer.

\end{itemize}

\begin{figure}
  \centering
    \includegraphics[width=0.45\textwidth]{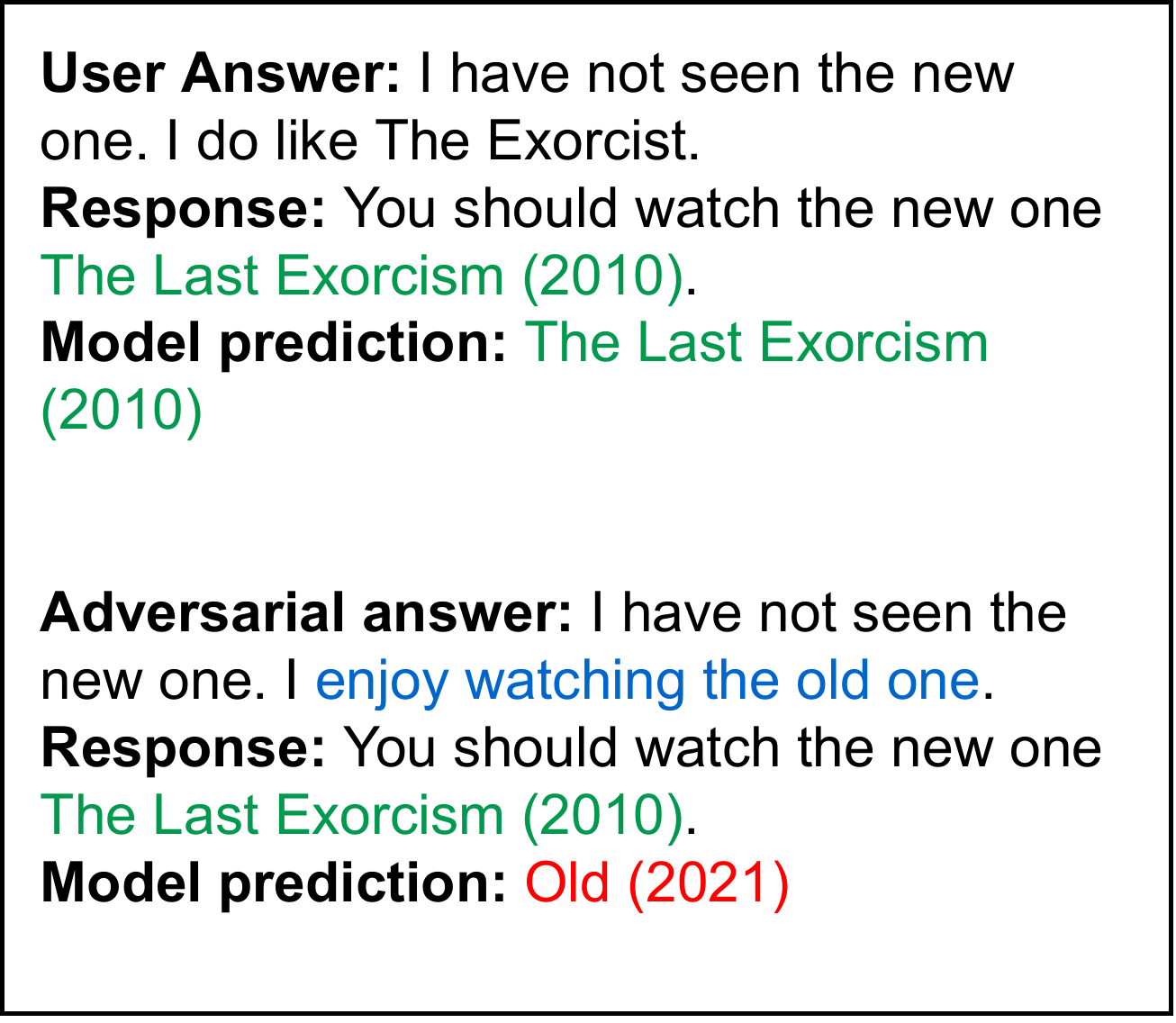}
    \caption{Expecting \textbf{same} prediction by \textbf{changing} user answer in the original conversation.}
    \label{fig:same_change}
\end{figure}

\begin{figure}
  \centering
    \includegraphics[width=0.45\textwidth]{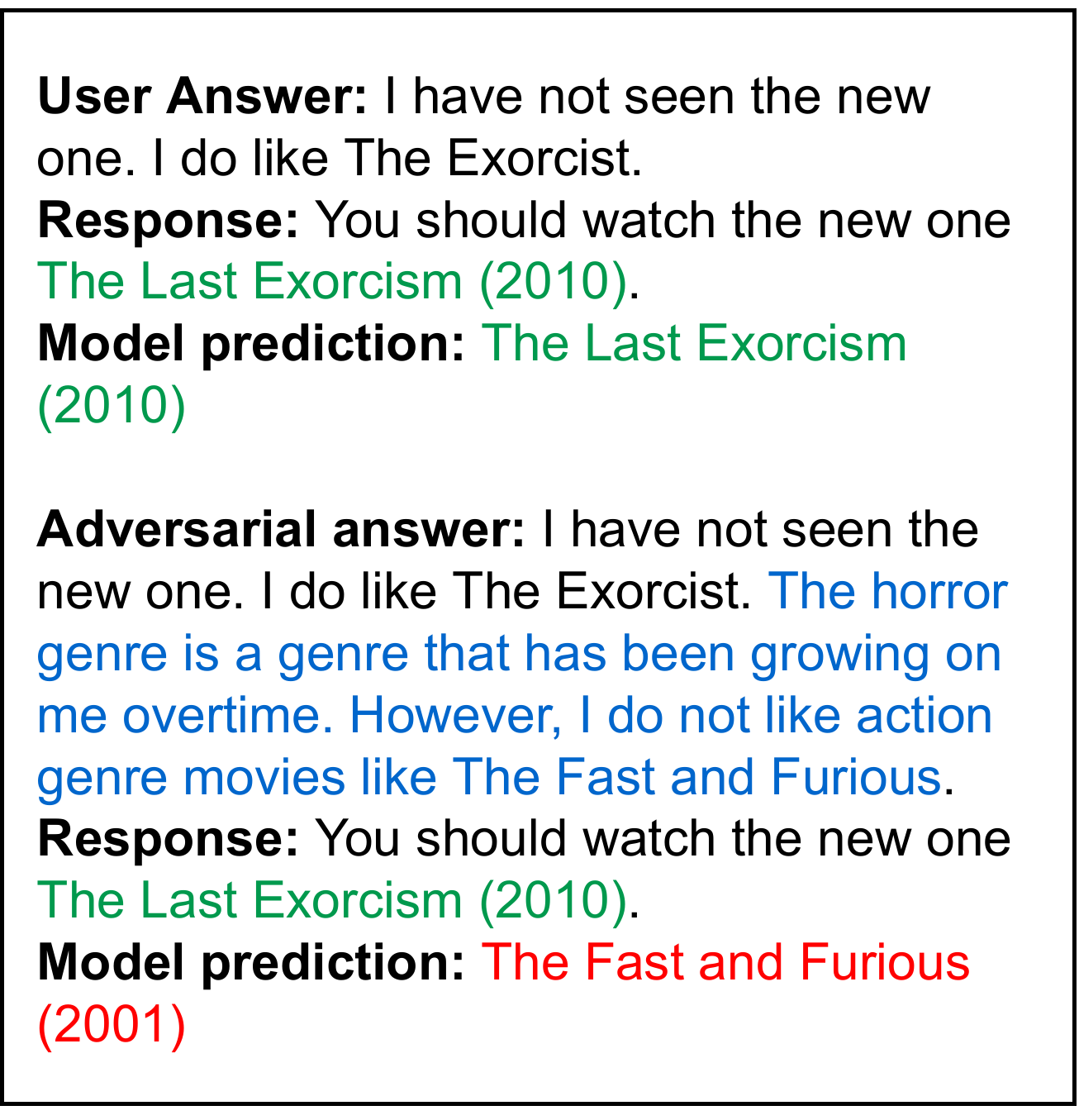}
    \caption{Expecting \textbf{same} prediction by \textbf{adding more details} to the user answer in the original conversation.}
    \label{fig:same_add}
\end{figure}

In the first category, by changing \emph{or} adding more details to the user's answer, we expect that the output of the system should not change. In the first scenario in this category, we change the user's answer in a way to make sure that we do not alter the semantic of the answer or user preferences. Therefore, a robust system should recommend the same item to the user in this case.
An example of this scenario is provided in Figure \ref{fig:same_change}.

In the second scenario in this category, we add some additional information in line with the user's response. This additional information should not change user preferences in the context of the conversation and simultaneously can distract the system.
We expect the same predictions by a robust CRS model when we add more details to the user's answer. This scenario ensures that the system is robust against long responses. We show an example of this scenario in Figure \ref{fig:same_add}.

In the second category, we aim to fool the system by altering the semantic of the conversation. Same as the first category, we design two scenarios for this category by changing \emph{or} adding a contradictory sentence to the user answer. However, in this category we expect the model output to be different than before. 

\begin{figure}
  \centering
    \includegraphics[width=0.45\textwidth]{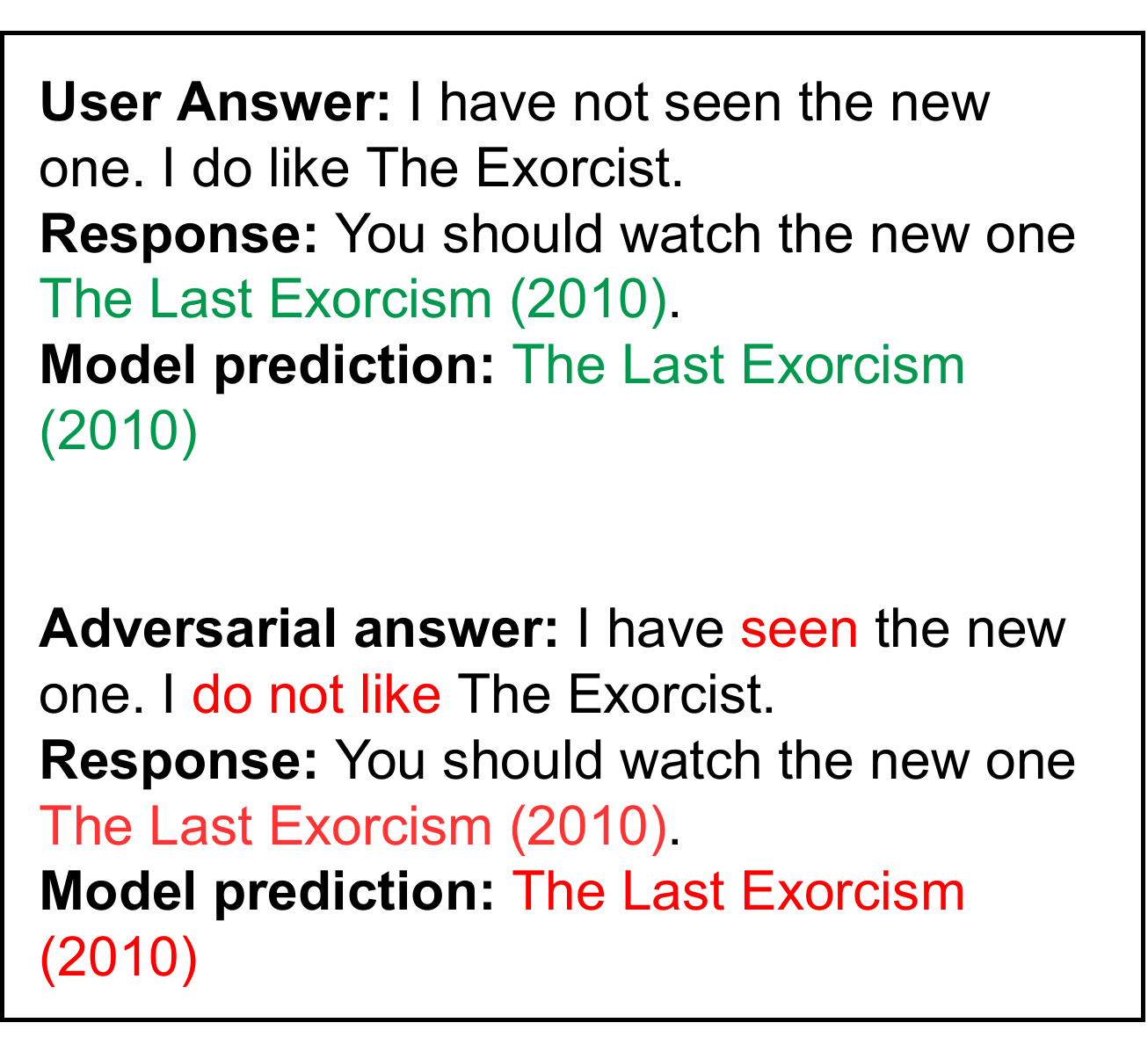}
    \caption{Expecting \textbf{different} prediction by \textbf{changing} user answer in the original conversation.}
    \label{fig:diff_change}
\end{figure}

In the first scenario of this category, we just change the user answer in a way that the semantic of it be different from before. 
The reason is that we want to make sure that the system output would be different compared to the recommendation for the not-attacked user's answer. If the system output is similar to before, this means the system is not able to satisfy the user preference. See an example of this scenario in Figure \ref{fig:diff_change}.

\begin{figure}
  \centering
    \includegraphics[width=0.45\textwidth]{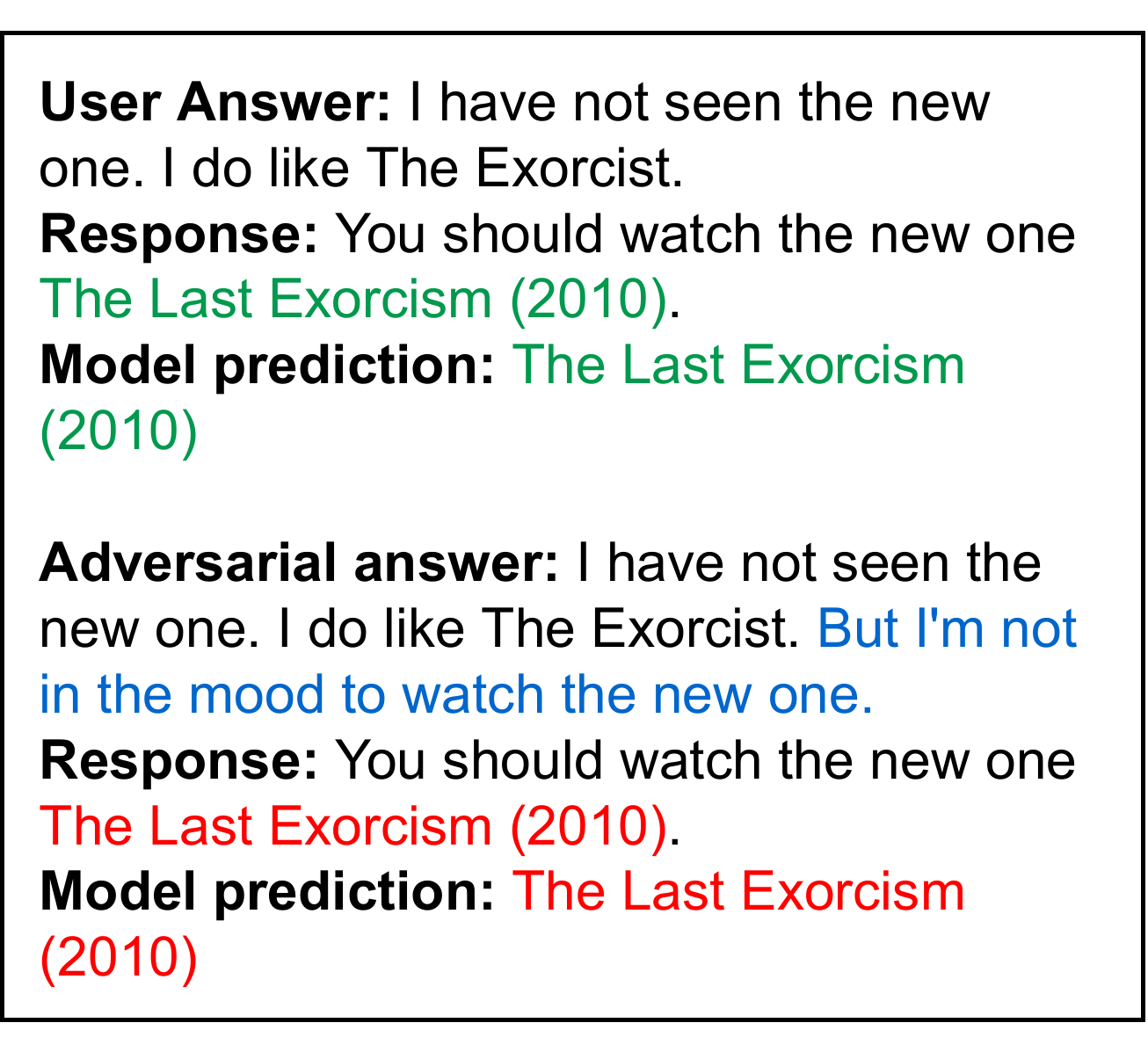}
    \caption{Expecting \textbf{different} prediction by \textbf{adding a contradictory sentence} to the user answer in the original conversation.}
    \label{fig:diff_add}
\end{figure}

In the second scenario, we just add a contradictory sentence to the user's answer and see the reaction of the system. By adding this contradictory sentence, we change the semantic of the user's answer and expect the model output would be different than before. An example of this scenario is shown in Figure \ref{fig:diff_add}.

In the following sections, we explain how to generate each of these four scenarios in two categories.

\subsection{Cat1: Expecting Same Prediction}
In this section, we explain how we generate each adversarial example in Cat1. In this category, we generate adversarial examples such that the meaning of the user's answer does not change. That's why we expect that the system responses do not change in this category.

\subsubsection{Changing the User's Answer by Semantic Preserving}
Given the user's answer, we aim to generate an alternative answer that looks similar to the user's answer but conveys the same meaning. 

To generate an answer, with the conditions mentioned above, we have several options including using synonyms and similar words, copying the idea and not the words, paraphrasing, using active and passive voice, etc. Since we want to make sure that the answer carries the same meaning, we only used synonyms and similar words because it would be difficult to generate them with other methods since it needs human labor which is time-consuming.
So, given a user's answer, we use a two-step procedure to generate these sentences. 

In the first step, we find the verb, the verb tense, and nouns of the sentence by Natural Language Toolkit (NLTK) \cite{loper2002nltk}. In the next step, we replace the verb and nouns with synonyms or similar words by NLTK. For example, given the user's answer ``I like watching horror movies'', we would like to change ``like'' to ``enjoy'' (a synonym word ``like''), ``horror'' to ``scary'', and ``movies'' to ``films'', resulting ``I enjoy watching scary films''. Note that we kept the verb tenses in the original sentence in the adversarial sentence. For each word, we try to find the most similar synonyms by NLTK. One can use other resources to find more effective synonyms and also try other methods to change the user's answer in this way.

\subsubsection{Adding More Details}
By adding more details, we can measure the system's ability to focus on the important parts of the user's response. So, we need to add additional information to the user's answer which this additional information should satisfy two constraints: 1) should not have any contradicts with the user's previous answers, and 2) should be also in line with the correct prediction. 

The generated more details depend on domains in datasets. There are two kind of domains, movies and books, in datasets that we use in this paper.

Therefore, given the context of the conversation, user's answer and correct prediction (i.e., movie or book), we generate this additional information by three steps:
\begin{itemize}
    \item first, we find the genre of the correct prediction by Wikipedia and DBpedia \cite{bizer2009dbpedia}, 
    \item in the next step, we just add the first sentence of Wikipedia of this genre to the user's answer. Note that this additional information should not change the model prediction,
    \item finally, we find a genre that is not mentioned in the conversation, and different from the correct genre and find a random movie from this genre. We add additional information to the user's answer to show that the user does not like this kind of movie.
\end{itemize}

For example, given that the user likes ``The Last Exorcism'', first we find the genre of this movie e.g., ``horror''. Then, we add the first sentence of Wikipedia of ``horror''. Finally, we find a random genre e.g., ``action'' and a random movie from this genre to distract a CRS model. In this example, we add ``The horror genre is a genre that has been growing on me overtime. However, I do not like action genre movies like \emph{The Fast and Furious}'' to the user's answer. Figure \ref{fig:same_add} illustrates an example of this category.

\subsection{Cat2: Expecting Different Prediction}
Contrary to the previous category, in this category, we aim to find adversarial examples such that the output of the model should be diff rent than before. In other words, we make sure that the user preferences are not compatible with the previous prediction of the model. Therefore, the model should change its prediction.
\subsubsection{Changing the User's Answer to the Opposite}
We want to make sure that the system can consider user preferences in the final results. So, we change the user's answer to the opposite one and expect that the model generates different output. 

To make an opposite sentence, we use two different methods: 1) finding the verb of the sentence and replacing it by its antonym, and 2) finding the verb of the sentence, recognizing the verb tense, and replacing it by ``not'', ``do not'', ``did not'', or ``does not''. As an example, given the user's answer, ``I like watching horror movies'', we would like to change ``like'' to ``hate'' (i.e., antonym of ``like'') or ``do not like''  and evaluate the output of the system. We find the verb and its tense by NLTK. Also, we use NLTK to find the antonym of a verb. One can use other resources to find more accurate antonyms. 

\subsubsection{Adding a Contradictory Sentence}
We also want to evaluate the ability of the system to deal with long sentences from the user, and also reasoning over the user's answer in this category.
So, we add a contradictory sentence to the user's answer and expect that the output generates different output compared to the original answer from the user. 
A contradictory sentence depends on the domain of the conversation. 
We used movies and books recommendation datasets in our experiments. Therefore, if we find a desire in the user for a movie or book in an answer, we change that desire by adding a contradictory sentence. For example, if the user's answer is ``I do like the Exorcist'', we add ``But I'm not in the mood to watch it'' to the user's answer. By adding this contradictory sentence, we want to make sure that the output of the model is not the same as before.

\section{Experiments}
For experiments, we use \emph{CRSLab} \cite{zhou2021crslab} which is an open-source toolkit for conversational recommender systems including several states of the art models and datasets. All three models are trained with Adam optimizer \cite{Kingma:2015} with $0.0001$ as the learning rate and $128$ as batch size. We trained all models with $50$ epochs to make sure that the models are trained well enough. Each utterance truncates after $256$ words to make sure that our additional information or contradictory sentences would be considered by models. Other parameters of each model are set based on suggested values in the original papers.

\subsection{Evaluation Measures}
For all experiments, we report hit@\{1, 10, 50\} \cite{yang2012top}, mrr@\{1, 10, 50\} \cite{vineela2021comprehensive}e, and ndcg@\{1, 10, 50\} \cite{jarvelin2002cumulated} which are used as recommendation metrics in \emph{CRSLab}.
\subsection{Datasets}
We use two commonly-used human-annotated CRS datasets 1) OpenDialKG \cite{moon2019opendialkg}, and 2) REDIAL \cite{li2018towards}.

OpenDialKG \cite{moon2019opendialkg} is an Open-ended Dialog with knowledge graph (KG) corpus which includes two different tasks: 1) Recommendation, and 2) Chit-chat. We use the first task in our work. OpenDialKG contains $13,802$ human-to-human  role-playing dialogs with $91,209$ utterances. For the recommendation task, we have movies and books domains in this dataset.

REDIAL is a REcommendations through DIALog dataset which is a conversational recommendation dataset collected by \cite{li2018towards}. REDIAL is constructed through Amazon Mechanical Turk (AMT). In seeker-recommender pairs, the AMT workers generate conversations for recommendations on movies.
It contains $10,006$ conversations consisting of $182,150$ utterances related to $51,699$ movies.

Table \ref{res:datasets} shows the basic statistics of these two datasets.
The datasets are split into training, validation, and test sets using a ratio of 8:1:1.

\subsection{CRS Models}
Our experiments are done on three CRS models: 1) ReDial \cite{li2018towards}, 2) KBRD \cite{chen2019towards}, and 3) KGSF \cite{zhou2020improving}.

\begin{itemize}
    \item ReDial \cite{li2018towards}: ReDial has been proposed with REDIAL dataset. The recommendation module is built on auto-encoder \cite{he2017distributed} and a sentiment analysis module.
    \item KBRD \cite{chen2019towards}: KBRD or Knowledge-Based Recommender Dialog system which is integration the recommender system and the dialog generation system. This model utilizes DBpedia \cite{bizer2009dbpedia} to enhance the semantics of contextual items or entities. 
    \item KGSF \cite{zhou2020improving}: KGSF is a novel KG-based semantic fusion approach for CRS. KGSF employs a KG-enhanced recommendation component for making accurate recommendations. To improve data representations in CRSs, KGSF uses both word-oriented and entity-oriented knowledge graphs (KG), as well as Mutual Information Maximization to align the word-level and entity-level semantic spaces. 
\end{itemize}
\begin{table}
\centering
\begin{tabular}{lcc}
\hline
\textbf{Dataset} & \textbf{\#Dialog} & \textbf{\#Utterance}\\
\hline
REDIAL & $10,006$ & $182,150$ \\
OpenDialKG & $13,802$ & $91,209$ \\
\hline
\end{tabular}
\caption{\label{res:datasets}
Basic statistics of the experimental datasets.
}
\end{table}

\begin{table*}
\centering
\scalebox{0.8}{
\begin{tabular}{clccccccccc}
\hline
\textbf{Model Name} & \textbf{Input Type} & \textbf{hit@1} & \textbf{hit@10} & \textbf{hit@50} & \textbf{mrr@1} & \textbf{mrr@10} & \textbf{mrr@50} & \textbf{ndcg@1} & \textbf{ndcg@10} & \textbf{ndcg@50}\\
\hline
\multirow{3}{*}{\textbf{ReDial}}
& Original & 0.0117 & 0.0367 & 0.1154 & 0.0117 & 0.0188 & 0.0225 & 0.0117 & 0.0230 & 0.0403 \\
 & Cat1-Change & 0.0016 & 0.0364 & 0.1085 & 0.0016 & 0.0087 & 0.0116 & 0.0016 & 0.0150 & 0.0301 \\
 & Cat1-Add & 0.0008 & 0.0355 & 0.1212 & 0.0008 & 0.0123 & 0.0161 & 0.0008 & 0.0179 & 0.0365 \\\hline
 \multirow{3}{*}{\textbf{KBRD}}
& Original & 0.2271 & 0.4466 & 0.5339 & 0.2271 & 0.3020 & 0.3057 & 0.2271 & 0.3370 & 0.3556  \\
 & Cat1-Change & 0.2144 & 0.4263 & 0.4890 & 0.2144 & 0.2875 & 0.2904 & 0.2144 & 0.3213 & 0.3349 \\ 
 & Cat1-Add & 0.1119 & 0.2669 & 0.3678 & 0.1119 & 0.1596 & 0.1646 & 0.1119 & 0.1852 & 0.2078\\ \hline
  \multirow{3}{*}{\textbf{KGSF}}
& Original & 0.2220 & 0.4331 & 0.5102 & 0.2220 & 0.2953 & 0.2994 & 0.2220 & 0.3287 &  0.3465\\
 & Cat1-Change & 0.2093 & 0.4229 & 0.5127 & 0.2093 & 0.2842 & 0.2890 & 0.2093 & 0.3180 & 0.3386 \\ 
 & Cat1-Add & 0.0864 & 0.2000 & 0.2975 & 0.0864 & 0.1211 & 0.1258 & .0864 & 0.1399 & 0.1615 \\ \midrule
\end{tabular}}
\caption{\label{res:cat1_opendialkg}
Adversarial evaluation by changing the user's answer or adding more details to it and expecting same prediction (Cat1), on OpenDialKG dataset.
}
\end{table*}

\begin{table*} 
\centering
\scalebox{0.8}{
\begin{tabular}{clccccccccc}
\hline
\textbf{Model Name} & \textbf{Input Type} & \textbf{hit@1} & \textbf{hit@10} & \textbf{hit@50} & \textbf{mrr@1} & \textbf{mrr@10} & \textbf{mrr@50} & \textbf{ndcg@1} & \textbf{ndcg@10} & \textbf{ndcg@50}\\
\hline
\multirow{3}{*}{\textbf{ReDial}}
& Original & 0.0127 & 0.0615 & 0.1811 & 0.0127 & 0.0240 & 0.0294 & 0.0127 &  0.0326 & 0.0586\\
 & Cat1-Change & 0.0107 & 0.0596 & 0.1773 & 0.0107 & 0.0224 & 0.0275 & 0.0107 & 0.0309 & 0.0561\\
 & Cat1-Add & 0.0107 & 0.0563 & 0.1781 & 0.0107 & 0.0227 & 0.0285 & 0.0107 & 0.0305 & 0.0575\\\hline
 \multirow{3}{*}{\textbf{KBRD}}
& Original & 0.0389 & 0.1825 & 0.3584 & 0.0389 & 0.0754 & 0.0839 & 0.0389 & 0.1004 & 0.1396\\
 & Cat1-Change & 0.0358 & 0.1850 & 0.3578 & 0.0358 & 0.0733 & 0.0816 & 0.0358 & 0.0994 & 0.1378  \\
 & Cat1-Add & 0.0331 & 0.1593 & 0.3170 & 0.0331 & 0.0643 & 0.0717 & 0.0331 & 0.0864 & 0.1212 \\\hline
  \multirow{3}{*}{\textbf{KGSF}}
& Original & 0.0347 & 0.1814 & 0.3716 & 0.0347 & 0.0720 & 0.0811 & 0.0347 & 0.0975 & 0.1398\\
 & Cat1-Change & 0.0345 & 0.1830 & 0.3733 & 0.0345 & 0.0720 & 0.0813 & 0.0345 & 0.0979 & 0.1403\\ 
 & Cat1-Add & 0.0289 & 0.1563 & 0.3255 & 0.0289 & 0.0602 & 0.0680 & 0.0289 & 0.0825 & 0.1197 \\\midrule
\end{tabular}}
\caption{\label{res:cat1_redial}
Adversarial evaluation by changing the user's answer or adding more details to it and expecting same prediction (Cat1), on REDIAL dataset.
}
\end{table*}

\subsection{Results and Discussion}
In this section, we evaluate three CRS models on two datasets for two categories.
\subsubsection{Adversarial Evaluation on Cat1}
 Results of these experiments are shown in Table \ref{res:cat1_opendialkg} and \ref{res:cat1_redial}  for OpenDialKG and REDIAL datasets, respectively.
Note that by adding adversarial examples in Cat1, we expect the same prediction by the models. Therefore, if by adding these examples to the user's answer the performance of a model drop, it would show the model is fooled by adversarial examples. The first observation is that KBRD and KGSF are much more effective approaches compared to the ReDial in the two datasets. Also, the performance of KBRD and KGSF on the OpenDialKG dataset are better than the REDIAL dataset. This can show that the REDIAL dataset is harder than the OpenDialKG dataset.

According to Table \ref{res:cat1_opendialkg} and \ref{res:cat1_redial}, both scenarios (i.e., changing the user's answer or adding more details to it) in two datasets can fool all CRS models. However, adding more details to the user's answer (i.e., Cat1-Add) is a more effective approach to fool the systems. 
Cat1-Add is even more successful in the OpenDialKG dataset and can reduce the performance of the systems up to $58 \%$. The reason is that by adding more details to the user's answer, CRS models are confused by additional information in the user's answer. Also, CRS models are fooled by changing the user's answer since we increase the difference between training data and test data. The reason is that CRS models can not properly use each word in the user's answer to find and recommend the appropriate item.
We believe that if a model needs to be successful in this task, should have a strong attention mechanism on the user's answer. This helps the system to fully understand user preferences. 

By showing the same results on two datasets, we verified that our adversarial examples are general enough to fool conversational recommender systems.

\begin{table*}
\centering
\scalebox{0.8}{
\begin{tabular}{clccccccccc}
\hline
\textbf{Model Name} & \textbf{Input Type} & \textbf{hit@1} & \textbf{hit@10} & \textbf{hit@50} & \textbf{mrr@1} & \textbf{mrr@10} & \textbf{mrr@50} & \textbf{ndcg@1} & \textbf{ndcg@10} & \textbf{ndcg@50}\\
\hline
\multirow{3}{*}{\textbf{ReDial}}
& Original & 0.0117 & 0.0367 & 0.1154 & 0.0117 & 0.0188 & 0.0225 & 0.0117 & 0.0230 & 0.0403 \\
 & Cat2-Change & 0.0101 & 0.0491 & 0.1119 & 0.0101 & 0.0177 & 0.0204 & 0.0101 & 0.0248 & 0.0383 \\
 & Cat2-Add & 0.0144 & 0.0381 & 0.0915 & 0.0144 & 0.0222 & 0.0247 & 0.0144 & 0.0260 & 0.0378 \\\hline
 \multirow{3}{*}{\textbf{KBRD}}
& Original & 0.2271 & 0.4466 & 0.5339 & 0.2271 & 0.3020 & 0.3057 & 0.2271 & 0.3370 & 0.3556 \\
 & Cat2-Change & 0.2178 & 0.4424 & 0.5195 & 0.2178 & 0.2939 & 0.2976 & 0.2178 & 0.3299 & 0.3470\\ 
 & Cat2-Add & 0.2407 & 0.4271 & 0.5068 & 0.2407 & 0.3030 & 0.3068 & 0.2407 & 0.3330 & 0.3508\\ \hline
  \multirow{3}{*}{\textbf{KGSF}}
& Original & 0.2220 & 0.4331 & 0.5102 & 0.2220 & 0.2953 & 0.2994 & 0.2220 & 0.3287 & 0.3465 \\
 & Cat2-Change & 0.2203 & 0.4331 & 0.5373 & 0.2203 & 0.2957 & 0.3005 & 0.2203 & 0.3292 & 0.3521\\ 
 & Cat2-Add & 0.2229 & 0.4356 & 0.5280 & 0.2229 & 0.2974 & 0.3021 & 0.2229 & 0.3310 & 0.3518 \\ \midrule
\end{tabular}}
\caption{\label{res:cat2_opendialkg}
Adversarial evaluation by changing the user's answer or adding a contradictory sentence to it and expecting different prediction (Cat2), on OpenDialKG dataset.
}
\end{table*}

\begin{table*}
\centering
\scalebox{0.8}{
\begin{tabular}{clccccccccc}
\hline
\textbf{Model Name} & \textbf{Input Type} & \textbf{hit@1} & \textbf{hit@10} & \textbf{hit@50} & \textbf{mrr@1} & \textbf{mrr@10} & \textbf{mrr@50} & \textbf{ndcg@1} & \textbf{ndcg@10} & \textbf{ndcg@50}\\
\hline
\multirow{3}{*}{\textbf{ReDial}}
& Original & 0.0127 & 0.0615 & 0.1811 & 0.0127 & 0.0240 & 0.0294 & 0.0127 & 0.0326 & 0.0586\\
 & Cat2-Change & 0.0127 & 0.0654 & 0.1797 & 0.0127 & 0.0234 & 0.0283 & 0.0127 & 0.0329 & 0.0573\\
 & Cat2-Add & 0.0127 & 0.0676 & 0.1886 & 0.0127 & 0.0253 & 0.0306 & 0.0127 & 0.0350 & 0.0611 \\\hline
 \multirow{3}{*}{\textbf{KBRD}}
& Original & 0.0389 & 0.1825 & 0.3584 & 0.0389 & 0.0754 & 0.0839 & 0.0389 & 0.1004 & 0.1396 \\
 & Cat2-Change & 0.0345 & 0.1875 & 0.3526 & 0.0345 & 0.0737 & 0.0815 & 0.0345 & 0.1003 & 0.1367 \\
 & Cat2-Add & 0.0358 & 0.1828 & 0.349 & 0.0358 & 0.0722 & 0.0802 & 0.0358 & 0.0978 & 0.1349\\\hline
  \multirow{3}{*}{\textbf{KGSF}}
& Original & 0.0347 & 0.1814 & 0.3716 & 0.0347 & 0.0720 & 0.0811 & 0.0347 & 0.0975 & 0.1398\\
 & Cat2-Change & 0.0358 & 0.1836 & 0.3791 & 0.0358 & 0.0724 & 0.0816 & 0.0358 & 0.0982 & 0.1414\\ 
 & Cat2-Add & 0.0345 & 0.1839 & 0.3763 & 0.0345 & 0.0709 & 0.0800 & 0.0971 & 0.0971 & 0.1397\\\midrule
\end{tabular}}
\caption{\label{res:cat2_redial}
Adversarial evaluation by changing the user's answer or adding a contradictory sentence to it and expecting different prediction (Cat2), on REDIAL dataset.
}
\end{table*}
\subsubsection{Adversarial Evaluation on Cat2}
We report the results of these experiments in Table  \ref{res:cat2_opendialkg} and \ref{res:cat2_redial} for  OpenDialKG and REDIAL datasets, respectively.
Note that in this category we expect different outputs by CRS models. In other words, the permanence of CRS models should not be the same. So, if the results are almost equal to the original results, this means that the system does not consider user preferences in the conversation at recommendations. If the performance of a CRS model drops significantly by seeing adversarial examples in Cat2, this can show the model is reliable.  

According to Table \ref{res:cat2_opendialkg} and \ref{res:cat2_redial}, in most cases, the results of all CRS models are comparable with two kinds of inputs i.e., original and adversarial inputs. In some cases, we can even see improvements over original inputs. We expect the results of CRS models to decrease significantly in face of adversarial examples in this category since by generating these adversarial examples, we make sure that the true recommendations are different from the ones that we have before. It can show that these CRS systems have many weaknesses in considering the user's preferences in the conversation. 
We hypothesize that a successful CRS model should satisfy two condition in the recommendation: 1) first needs to decide which words are important in the user's answer, 2) then consider all these important words (which are user preferences) in the recommendation. 
The ability of reasoning on the user's words can be very helpful for the development of these systems. For example,
for a user's answer, ``I'm not in the mood to watch it'', the system should be able to understand which movie ``it'' refers to in this sentence.
\citet{ma2021cr} emphasized that for accurate recommendation in this task, we need to have reasoning over the background knowledge and this is a big challenge in this area. \cite{fu2021hoops} is another approach for conversational recommendation in recent years which considers the importance of the reasoning for this task.

\section{Related Work}
\subsection{Conversational Recommender Systems}
Belkin et al. \cite{belkin1995cases} was a earliest work that proposed an interactive information retrieval system that utilized script-based conversational interaction for search. 
This area has gained much more popularity recently with the advent of
intelligent conversational systems and the process of neural approaches in the natural language process (NLP). Yang et al \cite{yang2017neural, yang2018response} proposed an approach to predict the next question in conversations.
A Multi-Memory Network (MMN) architecture for conversational search
and recommendation was proposed by Zhang et al. \cite{zhang2018towards}. 
A Belief Tracker model was developed by Sun et al. \cite{sun2018conversational}
to derive facet-value pairs from user utterances during the conversation. To decide between asking a question or recommending
an item, they also proposed a policy network.
Recently, Lei et al. \cite{lei2020estimation} showed that the interaction between conversation and recommendation can improve the performance of these systems substantially. They also proposed a policy network to decide between asking a question or recommending an item.
Zou et al. \cite{zou2020towards} proposed a question-based recommendation method that is able to ask users to express their preferences over descriptive item features. \citet{li2018towards} released a standard CRS dataset named REDIAL. They also proposed a hierarchical RNN model to generate utterances and recommendations. \citet{chen2019towards} and \citet{liao2019deep} use external Knowledge Graph (KG) to improve CRSs performance.
\citet{zhou2020towards} contributed a new CRS dataset named TG-ReDial and proposed the task of topic-guided conversational recommendation. They also proposed an effective approach for this task. \citet{moon2019opendialkg} proposed another dataset for the conversational recommendation. Each conversation turn is accompanied by a set of "KG routes" that connect the KG entities and relationships indicated in the dialog. Overall, these systems emphasize accurate recommendations, with simple or heuristic solutions implementing the dialogue component. 

\subsection{Adversarial Examples in NLP}
In NLP applications, algorithms have been developed to create adversarial sentences. \citet{papernot2016crafting} proposed an approach to generate adversarial examples on RNN/LSTM based model. 
\citet{li2016understanding} introduced an approach to learn the importance of words by deleting them in sentiment analysis task. Then, they use reinforcement learning to locate these words. 
By inserting and replacing words in a sentence with typos and synonyms, \citet{samanta2017towards} and  \citet{liang2017deep} proposed two approaches to generate adversarial sequences.
\citet{gao2018black} develop some scoring functions to find the most important words in the sentiment classification task and try to attack models in this task in a black-box setting.

\citet{jia2017adversarial} proposed several approaches to find distracting sentences and add them to a paragraph to fool a system in reading comprehension tasks. 
\citet{yang2020greedy} introduced a greedy approach to generate adversarial examples for discrete data by swapping the word or character.
\citet{ebrahimi2017hotflip} proposed an approach to generate adversarial examples on the character CNN model in the machine translation task. To specify which word should be replaced or deleted, they used the Jacobian matrix.
\citet{zhao2017generating} introduced an approach based on Generative Adversarial Networks (GANs) to generate natural adversarial examples.
\citet{cheng2020seq2sick} proposed some adversarial examples to evaluate the robustness of sequence-to-sequence models. They introduced a framework to generate non-overlapping and targeted keyword attacks on sequence-to-sequence models.
\citet{cheng2019evaluating} proposed algorithms to evaluate the robustness of a dialogue agent by designing adversarial agents. These agents are trained to attack a dialogue agent in both black-box and white-box settings.
\citet{ren2020crsal} proposed conversational recommender system with adversarial learning (CRSAL), a new end-to-end system to train a conversational recommendation system. \citet{cao2020adversarial} investigated adversarial attacks and 
detection approaches in reinforcement learning-based recommendation systems. They showed the effectiveness of these adversarial examples.

\section{Conclusions and Future Work}
We propose an adversarial evaluation scheme for the conversational recommendations task. The proposed scheme includes four scenarios in two categories. We show that how to generate these adversarial examples automatically. These adversarial examples can evaluate the robustness of conversational recommender systems. By evaluating three CRSs on two datasets, we show the effectiveness of these adversarial examples and the unreliability of these systems.
Our experimental results demonstrate the current CRSs are not robust against our adversarial examples.

Proposing and constructing more adversarial examples for this task can be interesting future work. In this paper, we just investigate adversarial examples for the evaluation of conversational recommender systems. However, it is possible to use these examples to iteratively train CRSs with them and improve the performance of these systems against adversarial examples. In this way, we can build more reliable CRSs to encourage people to use them.

\bibliography{anthology,custom}
\bibliographystyle{acl_natbib}

\appendix
\end{document}